# VirusT5: Harnessing Large Language Models to Predicting SARS-CoV-2 Evolution

Vishwajeet Marathe, Deewan Bajracharya and Changhui Yan

*Abstract*—During a virus's evolution, various regions of the genome are subjected to distinct levels of functional constraints. Combined with factors like codon bias and DNA repair efficiency, these constraints contribute to unique mutation patterns within the genome or a specific gene. In this project, we harnessed the power of Large Language Models (LLMs) to predict the evolution of SARS-CoV-2. By treating the mutation process from one generation to the next as a translation task, we trained a transformer model, called VirusT5, to capture the mutation patterns underlying SARS-CoV-2's evolution. We evaluated the VirusT5's ability to detect these mutation patterns including its ability to identify mutation hotspots and explored the potential of using VirusT5 to predict future virus strains. Our findings demonstrate the feasibility of using a large language model to model viral evolution as a translation process. This study establishes the groundbreaking concept of "mutation-as-translation," paving the way for new methodologies and tools in combating viral threats.

## I. INTRODUCTION

In the ongoing battle between humans and viruses, researchers have long aspired to predict how a virus might evolve, allowing them to design vaccines and prevention strategies in advance. The urgency of this goal reached unprecedented heights during the COVID-19 pandemic. Influenza serves as a prime example where significant efforts have been made to predict the dominant strain of the near future, enabling annual vaccine updates. A key element of this process is the global surveillance system, coordinated by the World Health Organization (WHO), which tracks and reports influenza cases over time and across different regions [1]. In parallel, substantial work is devoted to screening the antigenic properties of circulating strains, often using hemagglutinin inhibition (HAI) assays [2]. Some methods attempt to predict the evolution of the influenza virus by analyzing clade frequency trajectories and modeling the fitness function of a clade or lineage within the phylogenetic tree [3]. Other approaches focus on tracking mutations at protein sites critical to immune response and projecting the fitness landscape into the future [4].

Compared to Influenza, predictive models for SARS-CoV-2 are still in their early stages. This is not only due to the shorter history of study but also because of the unique challenges posed by the virus. Among these challenges is its high mutation rate [5,6] and higher genetic variability than initially anticipated [7]. It is estimated that the virus generates an average of 12 new effective RBD variants daily [5]. Other factors include the virus's ability to undergo recombination, the potential impact of population immunity from vaccination and prior infections on its evolution, and the statistical and computational hurdles in analyzing and interpreting the extensive data collected [8]. As the COVID pandemic emerged, a global surveillance system was rapidly established. Advances in sequencing technologies have allowed genomic sequences of SARS-CoV-2 isolates to be collected at an unprecedented pace. Today, more than 16 million genomic sequences are available for download in public repositories like GISAID [9]. Significant efforts have been made to analyze this data, enabling scientists to investigate outbreaks track the spread of the virus [10] and monitor the evolution of new variants [11].

Current predictive models for SARS-CoV-2 primarily focus on tracking mutations in the receptor binding domain (RBD) and other key regions of the spike protein that influence transmissibility, immune evasion, and vaccine efficacy. Wenkai Han et al. [12] combined structural modeling with deep learning to model antigenic evolution, demonstrating that their method could explore the viral fitness landscape and predict variants with enhanced immune evasion. Similarly, Nicole N. Thadani et al. [13] integrated fitness functions predicted by a deep learning model with biophysical and structural data to forecast the viral escape potential of variants. Bernadeta Dadonaite et al. [14] employed pseudovirus deep mutational scanning to assess how various mutations in the spike protein affect its binding affinity with ACE2, providing insights into potential future viral evolution. Soledad Delgado et al. [15] used self-organized maps to examine the diversification of haplotypes in infected patients, seeking potential strains that may emerge in the future.

Corresponding author: Changhui Yan

Vishwajeet Marathe is with the Department of Computer Science Department of North Dakota State University, Fargo, ND 58102 USA (e-mail: vishwajeet.marathe@ndsu.edu).

Deewan Bajracharya is with the Genomics, Phenomics and Bioinformatics Program of North Dakota State University, Fargo, ND 58102 USA (e-mail: vishwajeet.marathe@ndsu.edu).

Changhui Yan is with the Department of Computer Science and the Genomics, Phenomics and Bioinformatics Program of North Dakota State University, Fargo, ND 58102 USA (e-mail: changhui.yan@ndsu.edu).

Color versions of one or more of the figures in this article are available online at http://ieeexplore.ieee.org

This study specifically investigated viral evolution within a host rather than at the population level. While these advancements have deepened our understanding of how the virus may evolve, accurately predicting which strain will become the next dominant variant remains a significant challenge.

## II. DATASETS AND METHODS

### A. Genome Dataset.

This dataset comprises the genome sequences of 100,000 SARS-CoV-2 strains, serving as the corpus for pretraining our transformer model. To create this dataset, we downloaded complete, unambiguous (N%=0) SARS-CoV-2 genome sequences from the GISAID database [9]. From these, 100,000 sequences were randomly selected to reduce the required time for training. Pretraining utilized a maximum sequence length of 512 base pairs, so we divided the 100,000 genomes into non-overlapping segments of varying lengths, ranging from 1 to 512 base pairs.

### B. Receptor Binding Domain (RBD) Dataset

This dataset contains genetic sequences encoding the receptor-binding domain (RBD) and was used to fine-tune the pre-trained transformer model to build a predictor capable of classifying the RBD into four variant types. Additionally, this dataset was used to construct the Parent-Child dataset described in the following section. To obtain the dataset, we downloaded the codon-aware multiple sequence alignment (MSA) of SARS-CoV-2 genome sequences from GISAID. This MSA was subsequently aligned to the NCBI reference genome (NC_004718.3 SARS coronavirus Tor2) using the EBI's MUSCLE (Multiple Sequence Comparison by Log-Expectation) tool [16]. Finally, the RBD of each strain was extracted according to its coordinates in the reference genome.

### C. Parent-Child Dataset

This dataset contains pairs of RBD sequences, where the first sequence in each pair is considered the evolutionary parent of the second. We refer to these as parent-child pairs. This dataset was used to fine-tune the pre-trained transformer model to build a predictor capable of forecasting RBD sequence evolution.

To construct this dataset, we first used the GISAID metadata file to identify the variant type, lineage, and collection date for sequences in the RBD dataset. We focused on the four largest variant types—Alpha (49,234 sequences), Delta (173,300 sequences), Omicron (177,091 sequences), and non-VOC (77,063 sequences)—since the remaining variant types contained fewer than 5,000 sequences each. Together, these four variant types comprised 2,213 lineages with an average of 230 sequences per lineage.

Next, RBD sequences from the same lineage were divided into 10 consecutive bins based on their collection dates, ensuring each bin represented an equal time span. These bins, referred to as $B_1$, $B_2$, ..., $B_{10}$ in collection order, represent successive generations in the evolutionary process. Sequences in bin $B_i$ were considered evolutionary parents of sequences in bin $B_{i+1}$. We constructed 500,000 parent-child pairs by sampling sequences from these consecutive bins, with each pair consisting of one parental sequence from generation n and one child sequence from generation n+1.

### D. Pre-training of VirusT5

We utilized Google's "Text-to-Text Transfer Transformer" (T5) architecture for our VirusT5 model [17]. VirusT5 was pre-trained on the Genome Dataset using a masked language modeling (MLM) objective along with word dropout regularization, where 15% of the DNA sequences were masked and incorporated as sentinel tokens. We employed a maximum sequence length of 512 and a batch size of 100 segmented sequences.

To optimize training, we implemented an inverse square root learning rate schedule: the learning rate was set to 0.005 for the first 2,000 steps and then exponentially decayed until pretraining was complete. The model was pre-trained for a total of 12,000 steps.

### D. Fine-tuning of the pre-trained VirusT5

The fine-tuning dataset was randomly split into training, validation, and test sets with a 60:20:20 ratio. The pre-trained VirusT5 was fine-tuned on the training set and evaluated against the validation set at the end of each epoch. Fine-tuning continued until both the training and validation losses demonstrated minimal changes, indicating convergence. Finally, the fine-tuned VirusT5 model was tested on the test set to assess its performance.

### D. Computer system

VirusT5 was pre-trained and fine-tuned on high-performance computing (HPC) clusters at the Center for Computationally Assisted Science and Technology (CCAST) at North Dakota State University (NDSU). The computing setup included 32 CPU cores, 100 GB of RAM, and two NVIDIA A40 GPUs with 40 GB memory each.

## III. EXPERIMENTS AND RESULTS

### A. Classification of variant types based on the sequence of Receptor-Binding Domain (RBD)

We treated genomic sequences as a language, applying an NLP model to analyze the syntax and grammar of these sequences. To test the feasibility of this approach, we pre-trained the VirusT5 model on SARS-CoV-2 genome sequences and then fine-tuned it to predict the virus variant type using the DNA sequence of the receptor-binding domain (RBD) as input. The spike protein, responsible for facilitating the fusion of SARS-CoV-2 into host cells, initiates infection through the RBD, which recognizes and binds to host cell receptors. Mutations in the RBD affect the virus's infectiousness and virulence and serve as the primary criterion for classifying SARS-CoV-2 variants.

For the fine-tuning step, the RBD dataset was randomly divided into training, validation, and test sets with a 60:20:20 ratio. The pre-trained VirusT5 model was fine-tuned using the training set and evaluated on the validation set at the end of each epoch. As shown in Figure 1, both training and validation losses stabilized after 8 epochs, indicating convergence. Fine-tuning was therefore stopped after 8 epochs, and VirusT5 was then tested on the test set, achieving an accuracy of 97.29%. The confusion matrix of the predictions is shown in Figure 2. These results suggest that treating the SARS-CoV-2 genome as a type of natural language enables VirusT5 to learn the syntax and grammar of genomic sequences effectively, allowing it to accurately predict virus variants.

DNABERT-2 [18] is a recently published transformer-based method developed for a range of applications. In their study, the authors evaluated DNABERT-2's performance specifically in viral classification. For comparison, we downloaded the training, validation, and test datasets used by DNABERT-2 in their fine-tuning step. We then fine-tuned and tested our method on the same datasets. Our approach achieved an accuracy of 78.03% on the test dataset, outperforming the best version of DNABERT-2, which achieved an accuracy of 71.02%.

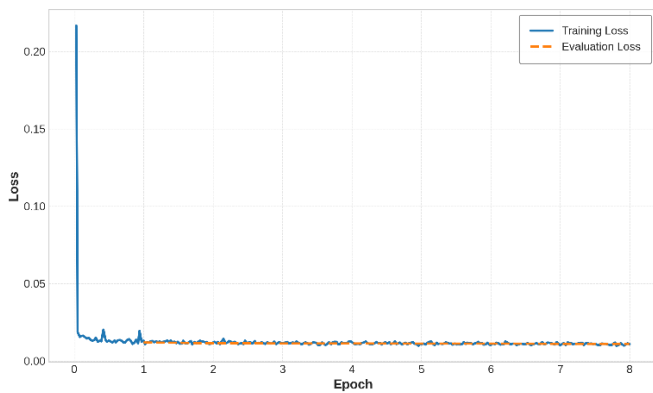

**Fig. 1.** Training loss and evaluation loss on the validation set stabilize after 8 epochs of fine-tuning.

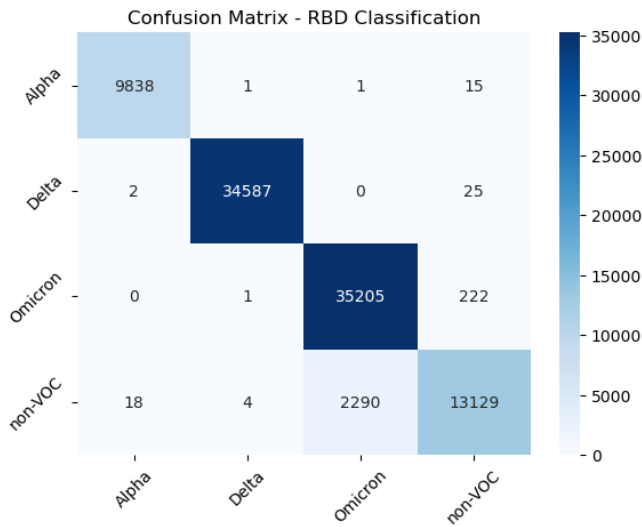

**Fig. 2.** Confusion matrix for the classification of variant types.

*B. Learning the evolutionary patterns using the transformer model*

SARS-CoV-2 mutates rapidly, but mutations in the spike protein gene are constrained by functionality and evolutionary pressures. In this task, we aim to test the feasibility of using VirusT5 to capture evolutionary patterns within the RBD of the spike gene and leverage the learned model to predict its future evolution. To approach this, we reformulate the gene mutation problem as a translation task. Let $S_n$ represent the RBD sequence of a SARS-CoV-2 strain from generation n. After accumulating mutations in one generation of time, Sn evolves into $S_{n+1}$. The process of Sn mutating into $S_{n+1}$ is analogous to translating Sn into Sn+1 using a language model that has learned the underlying mutation patterns We began by pretraining the VirusT5 model on the SARS-CoV-2 Genome Dataset, followed by fine-tuning with the Parent-Child Dataset. This dataset contained pairs of RBD sequences where the first sequence in each pair was the evolutionary parent of the second. The fine-tuning phase trained the model to "translate" a parental RBD sequence into its corresponding child sequence. We anticipated that through fine-tuning, the model would learn the evolutionary patterns specific to the RBD. To validate our approach, we conducted multiple evaluations of the model.

*B.1. Accuracy of translation*

We framed the evolutionary process as a translation task. For each parent-child pair, the model was trained to use the parental sequence as input and "translate" it into the child sequence. This approach equates to receiving an RBD sequence and predicting how it will evolve after one generation. The Parent-Child Dataset was divided into training, validation, and test sets using a 60:20:20 split. The pre-trained VirusT5 model was then fine-tuned on the training set and evaluated on the validation set at the end of each epoch. As shown in Figure 3, both training and validation losses stabilized with minimal change after 4 epochs. Consequently, we stopped fine-tuning after 4 epochs and tested the fine-tuned model on the test set.

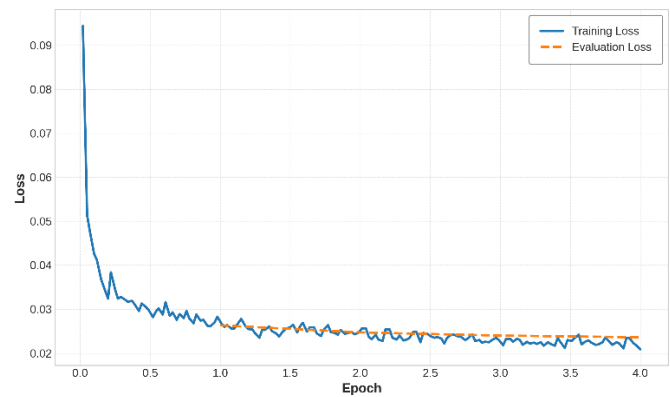

**Fig. 3.** Training loss and evaluation loss when the model was fine-tuned to translate a parental RBD sequence into a child sequence.

Translation accuracy was evaluated using the BLEU (Bilingual Evaluation Understudy) score and sequence identity. BLEU is an algorithm that assesses the quality of machine-translated text by comparing it to reference translations, with scores ranging from 0 (completely incorrect) to 1 (perfect translation). Our model achieved a BLEU score of 0.999 on the test set. The translation outputs were compared to the corresponding child sequences using a global alignment method, resulting in an average sequence identity of 99.97% with a standard deviation of 0.1%

These results validate the concept of transforming the viral evolution process into a translation problem, demonstrating that the transformer model successfully captured evolutionary patterns from the training set and utilized them to predict viral sequence evolution.

*B.2. Evolutionary Patterns That the Model Captured*

The RBD of the spike protein binds directly to the host receptor protein, with its functional stability being essential for viral infection and fusion. This requirement for functional stability imposes constraints on how the RBD can mutate, causing certain sites within the RBD to accumulate mutations more rapidly than others. These sites, known as mutation hotspots, are critical points in the mutation landscape.

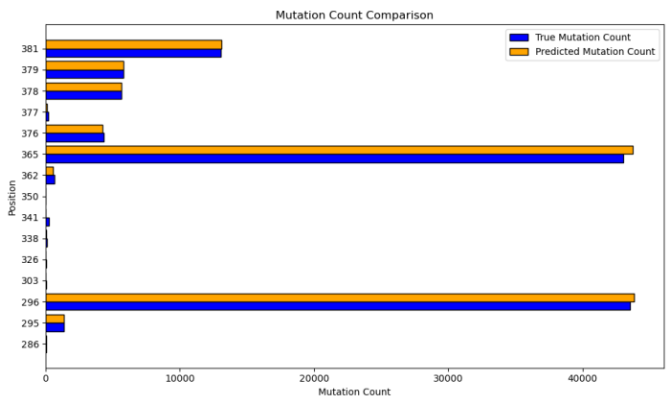

**Fig. 4.** Mutation pattern captured by VirusT5 (blue) align well with the rue mutation pattern (yellow).

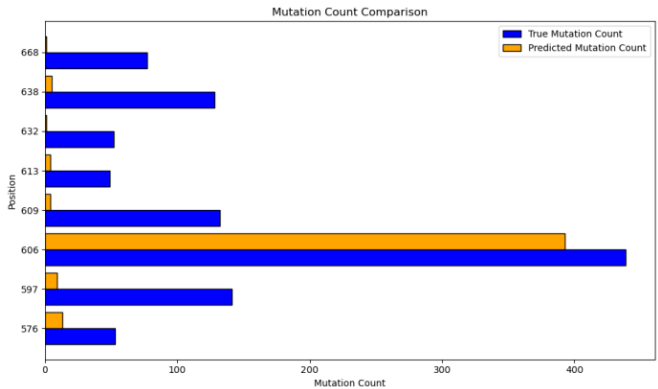

**Fig. 5.** Mutation pattern captured by VirusT5 (blue) slightly deviates from the rue mutation pattern (yellow) at the end of the RBD, but still accurately identifies the mutation hotspot (e.g. position 606)

To assess how well the fine-tuned VirusT5 captured these mutation patterns, we first aligned all child sequences with a reference RBD sequence and tallied the number of mutations at each reference site to represent the true observed mutation pattern. We then aligned all translation outputs (i.e., predicted sequences) to the reference sequence and counted the mutations at each reference site, reflecting the mutation patterns the model had learned.

The mutation patterns captured by VirusT5 showed a strong correlation with the true mutation patterns, with a Pearson correlation coefficient of 0.9999. Visual comparison revealed that VirusT5 accurately identified most mutation hotspots across the sequence, as shown in Figure 4, with the exception of the sequence's end region, detailed in Figure 5.

*C. A Generative Model for Virus Evolution.*

In Section B, we demonstrated that the VirusT5 model successfully learned evolutionary patterns and used them to predict RBD evolution over a single generation. Here, we extend this assessment to evaluate the model's ability to simulate the evolutionary process across multiple generations.

The parent-child dataset includes pairs from four major variant types: Alpha, Delta, Omicron, and non-VOC, with respective counts of 49,234, 173,300, 177,091, and 77,063 pairs. Following the protocol from Section 2, we fine-tuned four separate translation models, one for each variant type, denoted as $T_a$, $T_d$, $T_o$, and $T_n$ respectively. To simulate the evolutionary process, each model generated sequences iteratively as follows:

1. We used the Wuhan reference sequence as input to each model to generate a sequence of the 2nd generation.

2. This output was then used as the input to produce a 3rd-generation sequence.

3. This cycle was repeated until we obtained a sequence representing the 10th generation.

The 10th-generation sequence represents the present-day sequence, as sequences from the same lineage were divided into 10 generations based on collection time during data preparation. This process was repeated 100 times per model, generating 100 present-day sequences for each variant.

If the models accurately simulate the natural evolutionary process, sequences generated by each model should reflect the properties of their corresponding variant type. For example, the sequences generated by model $T_a$, should exhibit characteristics of the Alpha variant. To test this, we fed the 400 present-day sequences generated by the four models into the classifier trained in Section 1 to predict their variant types. The classifier achieved 100% accuracy in variant classification, as shown in Figure 6's confusion matrix.

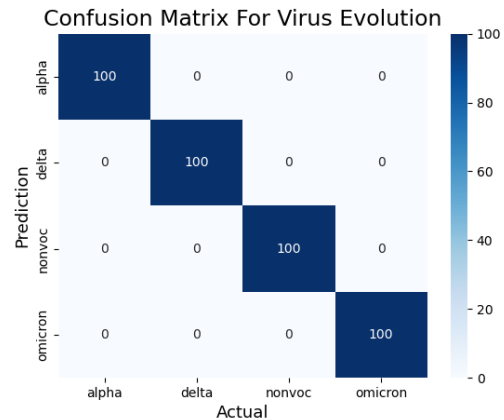

**Fig. 6.** Confusion matrix for classification of model-

generated sequences

It is noteworthy that all present-day sequences were generated from the same Wuhan reference sequence. Each sequence underwent a distinct evolutionary process depending on the translation model it was passed through. The classification results confirm that the fine-tuned translation models can be applied iteratively to simulate the evolutionary process, effectively predicting how the virus may mutate along different lineages.

## IV. SUMMARY AND DISCUSSION

Viral evolution is a complex process influenced by numerous factors, including mutations and immunization efforts. The rapid mutation rate of the SARS-CoV-2 virus presents a challenge to the effectiveness of existing vaccines and therapies. Anticipating potential future mutations could enable preemptive vaccine development. In this paper, we propose a novel approach that models viral evolution as a language translation process. We trained a transformer-based language model to capture the specific mutation patterns within the receptor-binding domain (RBD) of SARS-CoV-2. We evaluated our model's performance based on its translation accuracy and ability to identify mutation hotspots. Additionally, we assessed the feasibility of using this model as a generative tool to forecast future virus strains. Our findings confirm that the transformer-based language model successfully learns unique viral mutation patterns and generates sequences for potential future strains.

This study establishes the groundbreaking concept of "mutation-as-translation," paving the way for new methodologies and tools in combating viral threats. While this approach has proven effective, certain limitations remain. First, the virus strains were categorized into ten bins, each representing a generation of the virus; however, in reality, the temporal intervals of these bins may not align with actual viral generation times. Due to the lack of ground-truth data identifying the generational stage of each strain, these estimates remain approximate. Second, our data consisted of Parent-Child pairs within lineages, which reflect population-level lineage evolutions across various geographical regions rather than true biological parent-child relationships. Obtaining data that accurately reflects a biological parent-child relationship would require tracking viral strains within individual hosts, as demonstrated in studies like [15]. Unfortunately, data of this nature are rare and currently insufficient for training large language models at the necessary scale.

## V. DATASET AND CODE AVAILABILITY

The model, code are available for use at https://github.com/vrmarathe/VirusT5 .The pretraining dataset, finetuning dataset are available on request.

generated sequences

It is noteworthy that all present-day sequences were generated from the same Wuhan reference sequence. Each sequence underwent a distinct evolutionary process depending on the translation model it was passed through. The classification results confirm that the fine-tuned translation models can be applied iteratively to simulate the evolutionary process, effectively predicting how the virus may mutate along different lineages.

## IV. SUMMARY AND DISCUSSION

Viral evolution is a complex process influenced by numerous factors, including mutations and immunization efforts. The rapid mutation rate of the SARS-CoV-2 virus presents a challenge to the effectiveness of existing vaccines and therapies. Anticipating potential future mutations could enable preemptive vaccine development. In this paper, we propose a novel approach that models viral evolution as a language translation process. We trained a transformer-based language model to capture the specific mutation patterns within the receptor-binding domain (RBD) of SARS-CoV-2. We evaluated our model's performance based on its translation accuracy and ability to identify mutation hotspots. Additionally, we assessed the feasibility of using this model as a generative tool to forecast future virus strains. Our findings confirm that the transformer-based language model successfully learns unique viral mutation patterns and generates sequences for potential future strains.

This study establishes the groundbreaking concept of "mutation-as-translation," paving the way for new methodologies and tools in combating viral threats. While this approach has proven effective, certain limitations remain. First, the virus strains were categorized into ten bins, each representing a generation of the virus; however, in reality, the temporal intervals of these bins may not align with actual viral generation times. Due to the lack of ground-truth data identifying the generational stage of each strain, these estimates remain approximate. Second, our data consisted of Parent-Child pairs within lineages, which reflect population-level lineage evolutions across various geographical regions rather than true biological parent-child relationships. Obtaining data that accurately reflects a biological parent-child relationship would require tracking viral strains within individual hosts, as demonstrated in studies like [15]. Unfortunately, data of this nature are rare and currently insufficient for training large language models at the necessary scale.

## V. DATASET AND CODE AVAILABILITY

The model, code are available for use at https://github.com/vrmarathe/VirusT5 .The pretraining dataset, finetuning dataset are available on request.

## REFERENCES

[1] Ampofo, W. K. et al. Improving influenza vaccine virus selection: report of a WHO informal consultation held at WHO headquarters, Geneva, Switzerland, 14-16 June 2010. Influenza Other Respiratory Viruses, 6, 142–152, 2012.

[2] Smith, D. J. et al. Mapping the antigenic and genetic evolution of influenza virus. Science, 305, 371–376, 2004).

[3] Łuksza, M. & Lässig, M. A predictive fitness model for influenza. Nature, 507, 57–61, 2014; Huddleston, J. et al. Integrating genotypes and phenotypes improves long-term forecasts of seasonal influenza A/H3N2 evolution. eLife, 9, e60067, 2020; Neher, R. A., Russell, C. A. & Shraiman, B. I. Predicting evolution from the shape of genealogical trees. eLife, 3, e03568, 2014

[4] Lou, J., Liang, W., Cao, L. et al. Predictive evolutionary modelling for influenza virus by site-based dynamics of mutations. Nat Commun 15, 2546 (2024)

[5] Duarte, C.M., Ketcheson, D.I., Eguíluz, V.M. et al. Rapid evolution of SARS-CoV-2 challenges human defenses. Sci Rep 12, 6457 (2022)

[6] Markov, P.V., Ghafari, M., Beer, M. et al. The evolution of SARS-CoV-2. Nat Rev Microbiol 21, 361–379 (2023)

[7] H. A. Al Khatib et al., Within-host diversity of SARS-CoV-2 in COVID-19 patients with variable disease severities. Front. Cell Infect. Microbiol. 10, 575613 (2020).

[8] Cappello L, Kim J, Liu S, Palacios JA. Statistical Challenges in Tracking the Evolution of SARS-CoV-2. Stat Sci. 2022 May;37(2):162-182.

[9] Shu Y, McCauley J. GISAID: Global initiative on sharing all influenza data - from vision to reality. Euro Surveill. 2017 Mar 30;22(13):30494.

[10] Deng X, Gu W, Federman S, Du Plessis L, Pybus OG, Faria NR, Wang C, Yu G, Bushnell B et al. (2020) Genomic surveillance reveals multiple introductions of Sars-CoV-2 into northern California. Science 369 582–587.

[11] Volz E, Hill V, McCrone JT, Price A, Jorgensen D, O'toole Á, Southgate J, Johnson R, Jackson B et al. (2021). Evaluating the effects of Sars-CoV-2 spike mutation D614G on transmissibility and pathogenicity. Cell 184 64–75.

[12] Han, W., Chen, N., Xu, X. et al. Predicting the antigenic evolution of SARS-COV-2 with deep learning. Nat Commun 14, 3478 (2023)

[13] Thadani, N.N., Gurev, S., Notin, P. et al. Learning from prepandemic data to forecast viral escape. Nature 622, 818–825 (2023).

[14] Dadonaite, B., Brown, J., McMahon, T.E. et al. Spike deep mutational scanning helps predict success of SARS-CoV-2 clades. Nature 631, 617–626 (2024).

[15] Delgado S, Somovilla P, Ferrer-Orta C, Martínez-González B, Vázquez-Monteagudo S, Muñoz-Flores J, Soria ME, García-Crespo C, de Ávila AI, Durán-Pastor A, Gadea I, López-Galíndez C, Moran F, Lorenzo-Redondo R, Verdaguer N, Perales C, Domingo E. Incipient functional SARS-CoV-2 diversification identified through neural

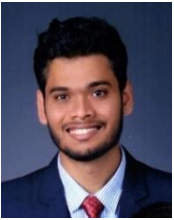

**Vishwajeet Marathe** is a PhD candidate in the Departmentof Computer Science at North Dakota State University (NDSU) in Fargo, North Dakota, USA. His research focuses on the application of machine learning, deep learning and large language models to bioinformatics.

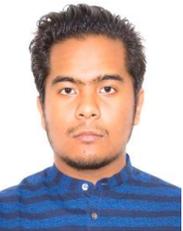

**Deewan Bajracharya** is a PhD candidate in the Genomics, Phenomics and Bioinformatics Program at North Dakota State University (NDSU) in Fargo, North Dakota, USA. His research interests and field of interest include bioinformatics, viral genomics, and transcriptomics where he leverages bioinformatics tools and pipelines to study host-pathogen interactions, host immune responses, and pathogen/variant detection.

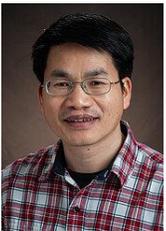

**Changhui Yan** is a Professor in the Department of Computer Science at North Dakota State University (NDSU) in Fargo, North Dakota, USA. He also serves as the Director of NDSU's Genomics, Phenomics and Bioinformatics Program. His research interests encompass a diverse range of fields, including machine learning, bioinformatics, computational biology, genomics, cloud computing, and data science.